\documentclass[sigconf]{acmart}

\AtBeginDocument{%
  \providecommand\BibTeX{{%
    \normalfont B\kern-0.5em{\scshape i\kern-0.25em b}\kern-0.8em\TeX}}}

\copyrightyear{2024} 
\acmYear{2024} 
\setcopyright{acmlicensed}\acmConference[Websci '24]{ACM Web Science Conference}{May 21--24, 2024}{Stuttgart, Germany}
\acmBooktitle{ACM Web Science Conference (Websci '24), May 21--24, 2024, Stuttgart, Germany}
\acmDOI{10.1145/3614419.3644026}
\acmISBN{979-8-4007-0334-8/24/05}

\usepackage[justification=centering,font=small]{caption,subcaption}
\usepackage{cleveref}
\usepackage{xcolor}
\usepackage{booktabs}
\usepackage{multirow} 
\usepackage{tikz}
\usetikzlibrary{calc}
\usepackage[textwidth=\marginparwidth, textsize=tiny, color=green!30]{todonotes}
\definecolor{betteryellow}{RGB}{237, 232, 90}

\begin{document}

\title{Deciphering Crypto Twitter} 

\author{Inwon Kang}
\affiliation{%
  \institution{Rensselaer Polytechnic Institute}
  \streetaddress{110 8th Street}
  \city{Troy}
  \state{New York}
  \country{USA}
  \postcode{12180}
}
\email{kangi@rpi.edu}
\orcid{0000-0001-8912-287X}

\author{Maruf Ahmed Mridul}
\affiliation{%
  \institution{Rensselaer Polytechnic Institute}
  \streetaddress{110 8th Street}
  \city{Troy}
  \state{New York}
  \country{USA}
  \postcode{12180}
}
\email{mridum@rpi.edu}
\orcid{0009-0003-7501-4714}

\author{Abraham Sanders}
\affiliation{%
  \institution{Rensselaer Polytechnic Institute}
  \streetaddress{110 8th Street}
  \city{Troy}
  \state{New York}
  \country{USA}
  \postcode{12180}
}
\email{sandea5@rpi.edu}
\orcid{0000-0002-5231-2239}

\author{Yao Ma}
\affiliation{%
  \institution{Rensselaer Polytechnic Institute}
  \streetaddress{110 8th Street}
  \city{Troy}
  \state{New York}
  \country{USA}
  \postcode{12180}
}
\email{may13@rpi.edu}
\orcid{0000-0002-4985-8724}

\author{Thilanka Munasinghe}
\affiliation{%
  \institution{Rensselaer Polytechnic Institute}
  \streetaddress{110 8th Street}
  \city{Troy}
  \state{New York}
  \country{USA}
  \postcode{12180}
}
\email{munast@rpi.edu}
\orcid{0000-0002-0911-750X}

\author{Aparna Gupta}
\affiliation{%
  \institution{Rensselaer Polytechnic Institute}
  \streetaddress{110 8th Street}
  \city{Troy}
  \state{New York}
  \country{USA}
  \postcode{12180}
}
\email{guptaa@rpi.edu}
\orcid{0000-0002-5275-7756}

\author{Oshani Seneviratne}
\affiliation{%
  \institution{Rensselaer Polytechnic Institute}
  \streetaddress{110 8th Street}
  \city{Troy}
  \state{New York}
  \country{USA}
  \postcode{12180}
}
\email{senevo@rpi.edu}
\orcid{0000-0001-8518-917X}

\renewcommand{\shortauthors}{Kang et al.}

\begin{abstract}
Cryptocurrency is a fast-moving space, with a continuous influx of new projects every year.
However, an increasing number of incidents in the space, such as hacks and security breaches, threaten the growth of the community and the development of technology.
This dynamic and often tumultuous landscape is vividly mirrored and shaped by discussions within ``Crypto Twitter,'' a key digital arena where investors, enthusiasts, and skeptics converge, revealing real-time sentiments and trends through social media interactions.
We present our analysis on a Twitter dataset collected during a formative period of the cryptocurrency landscape. 
We collected 40 million tweets using keywords related to cryptocurrency and performed a nuanced analysis that involved grouping the tweets by semantic similarity and constructing a tweet and user network. 
We used sentence-level embeddings and autoencoders to create K-means clusters of tweets.
We identified six groups of tweets and their topics to examine different cryptocurrency-related interests and the change in sentiment over time.
For example, we identified different groups of tweets demonstrating coordinated behavior in the market or expressing distrust in centralized cryptocurrency exchanges.
Moreover, we discovered sentiment indicators that point to real-life incidents in the crypto world, such as the FTX incident of November 2022.
We also constructed and analyzed different networks of tweets and users in our dataset by considering the reply and quote relationships and analyzed the largest components of each network.
Our networks reveal a structure of bot activity in Crypto Twitter and suggest that they can be detected and handled using a network-based approach.
Our work sheds light on the potential of social media signals to detect and understand crypto events, benefiting investors, regulators, and curious observers alike, as well as the potential for bot detection in Crypto Twitter using a network-based approach. 

\end{abstract}



\maketitle
\section{Introduction}
The dynamic and often unpredictable nature of the cryptocurrency market, where rapid innovation intersects with complex security challenges, has rendered it a unique domain for exploring the interplay between technological developments and public sentiment, particularly as mirrored in the vast digital landscape of social media.
Following the release of Ethereum in 2015, the crypto world has gained access to ever-increasing programmability in the blockchain, leading to the \textit{Initial Coin Offerings} (ICO) craze of 2017 \cite{fenu2018ico,gupta2020descriptive}.
While that excitement has toned down,  a steady stream of new projects in the cryptocurrency world are constantly ``advertised'' and discussed on social media.

However, recent years have seen an increasing number of attacks on blockchain projects~\cite{chen2020survey}, much of those stemming from social engineering attacks that originate on social media.
CNBC reports that nearly \$4 billion were stolen in just 2022 \cite{cnbc2023cryptoloss}, and the number of lawsuits related to these losses continuously increased, reaching up to 20 suits filed annually by 2023~\cite{bloomberg2023cryptolawsuit}.
In one instance, Axie Infinity, a platform that promised users real-life returns for playing, i.e., a ``play-to-earn'' game, suffered a devastating hack that lost over \$620 million:
40\% of the platform's users were from the Philippines, many of whom had invested their life savings into playing the game in hopes of earning high returns from the platform~\cite{Chow2022Jul}.
The hack left many users in crippling debt with little to no reparations~\cite{Kaaru2022Jul}. 
This Ronin Bridge hack of Axie Infinity, one of the largest security breaches in the crypto space, not only had a profound impact on the game's ecosystem but also sparked extensive discussions across social media platforms, exemplifying how significant real-world events in the cryptocurrency world can rapidly translate into widespread conversations and sentiment shifts in the digital realm.

As the communities behind various blockchain projects grow, there is increasing activity on social media related to these topics.
These discussions can offer valuable insights on various topics in the blockchain world and possibly even forecast events that take place in real life.
However, the large volume of online activity also incurs difficulties in finding valuable information due to the varying quality of contributions and the sheer volume of the data.
We addressed this issue by leveraging semantic-aware embeddings and a network-based approach to parse the data.

Social media platforms such as Twitter (now called X) allow for many types of analysis thanks to the textual nature of the posts.
The tweets can be used to detect the topic of different groups of users, as well as approximate the sentiment across different topics using pre-trained language models such as BERT~\cite{devlin2018bert} or RoBERTa~\cite{liu2019roberta}.
In some cases, these insights can provide signals pertaining to certain events' occurrence~\cite{garcia2018sentiment}.
Similarly, the sentiment analysis of tweets related to blockchain technologies and decentralized finance (DeFi) can be leveraged to understand the broad user base sentiment towards various projects or aspects of blockchain technologies.

In this work, we conduct a large-scale analysis of Twitter posts related to cryptocurrencies collected to understand the possible correlations between signals in social media and real-life events and the conversation patterns found in Crypto Twitter\footnote{Crypto Twitter refers to the vibrant and active online community on the Twitter platform, where enthusiasts, investors, developers, and thought leaders in the cryptocurrency space congregate to share news, opinions, analyses, and insights about various aspects of cryptocurrencies and blockchain technology.}
We used Twitter's v1 API \cite{twitterapi} to collect tweets that contain cryptocurrency-related keywords and built a large dataset of around 40 million tweets.
Our data was collected immediately following the height of cryptocurrency bridge hack incidents between November 9, 2022, and November 23, 2022.
It is also worth noting that during this period, the FTX bankruptcy was made public~\cite{Sigalos2022Nov}, which was a formative period in public sentiment toward cryptocurrency.
Thus, we find many mentions related to this incident in our dataset.
However, instead of focusing on a specific attack vector or an incident, we aim to do a broader analysis of real-life events in cryptocurrency and possible signals that can be detected from social media.
We achieve this by conducting an overall analysis of Crypto Twitter, scanning the broad sphere to find correlations between various real-life events and the users' reactions.

\noindent Our contributions are as follows:
\begin{itemize}
    \item Topic modeling and sentiment analysis of tweets specific to blockchain projects or issues.
    \item A study of the correlation between sentiment scores of different tweet groups and real-life blockchain hacks, scams, and incidents.
    \item A graph-based analysis of tweet-level and user-level interactions in Crypto Twitter to further understand the social dynamics of the space.
\end{itemize}

\section{Related Work}
In navigating the intricate landscape of cryptocurrency and its social media dynamics, it is essential to explore the existing body of research, which sheds light on the multifaceted interactions between digital currency markets and online communities and the broader implications of these relationships on market sentiment and user behavior.
Building upon this foundation, \citet{guggenberger2021} have significantly contributed to the literature on blockchain security, noting a marked increase in research in this area since 2013. Complementing this, \citet{vokerla2019overview} discusses a more focused analysis, characterizing ten prevalent methods of exploits across five distinct applications of blockchain technology.

\citet{mirtaheri2019} used a random forest classifier to predict occurrences of cryptocurrency pump-and-dump scams using social media signals.
Since pump-and-dump scams encourage the purchase of assets to inflate prices artificially, the authors posit that sudden increases in chatter surrounding a particular project may indicate a pump-and-dump scheme in progress.
The resulting classifier predicts these schemes across social media platforms, such as Twitter and Telegram, with an average accuracy of 74\% $\pm$ 8\%.
The authors also identify an increase in bot accounts for generating artificial posts while a scheme is underway.
However, unlike \citet{mirtaheri2019}, which only focused on pump-and-dump schemes, we analyze various security flaws, hacks, and sentiments related to major events in the crypto space.

\citet{Saad2020exploring} analyzed the frequency of Google searches using terms related to a type of attack known as \textit{cryptojacking}.
The findings reveal a positive correlation between search term frequency and a large-scale attack of the indicated character.
The authors discuss attack vectors in theory and in relation to high-profile attacks such as the one that bankrupted the cryptocurrency exchange Mt. Gox in 2013~\cite{cheung2015crypto}. 

Previous work has also used social media data to find the correlation between user sentiments and the price of different cryptocurrencies.
\citet{huang2021lstm} applied a long-short term memory (LSTM) based model to analyze Chinese social media sentiment and accurately predict cryptocurrency price fluctuations.
The model used a lexicographical approach to creating a custom dictionary of cryptocurrency terms and achieved performance increases in precision and recall over a standard auto-regression model for time-series prediction.
\citet{pano2020complete} used tweets related to Bitcoin to calculate the VADER sentiment scores~\cite{hutto2014vader} and obtained the score's correlation with Bitcoin price data.
\citet{lamon2017cryptocurrency} leveraged Twitter sentiment analysis to build a price prediction model for Bitcoin, Ethereum, and Litecoin.
The authors find that the model can successfully predict large margins of price change, although more specific predictions are more difficult.
\citet{kraaijeveld2020predictive} employed a similar analysis using Twitter data against cryptocurrency prices, concluding that Twitter data are a good predictor for some cryptocurrencies, such as Bitcoin, Ethereum, and Litecoin.
The authors also note the heavy presence of bot accounts in the blockchain Twitter space, noting that 1$\sim$14\% of the collected data are likely spam. 


Finally, the work most similar to ours is presented by \citet{Linton2017} and \citet{parkHowAreTwitter2019}.

\citet{Linton2017} used a modified form of Latent Dirichlet Allocation (LDA) capable of preserving time-series relationships in their study of topic modeling of discussion forum posts retrieved from bitcointalk.org, a platform for discussing cryptocurrencies and crypto-related events.
Their findings, of particular interest for our experiments, support the identification of cryptocurrency-related events like suspected attacks on exchanges through clustering.
It should be noted that their work finds that chatter related to all attacks was joined in a single cluster rather than one per attack, suggesting the need for further research into models that enhance attack-level granularity.
In our work, we consider Twitter as the data source and use topic modeling and sentiment analysis to draw correlations to real-life events.

\citet{parkHowAreTwitter2019} analyze different types of networks in Crypto Twitter to identify the correlation in project ratings of different cryptocurrencies.
Specifically, the authors use different relationships between tweets, such as follow-follow, reply-mention, and quote, of the official accounts of cryptocurrency projects to construct networks and draw a correlation to the Weiss Rating, which evaluates the financial strength and investment risk of various cryptocurrency projects aimed at guiding investors with independent and unbiased evaluation of the potential and stability of these projects.
The authors find that the follow-follow network correlates highly with the Weiss Rating.
However, this work focuses on the network structure of only \textit{official accounts}, while our work considers the Twitter networks of \textit{all} users we have in the dataset.

\section{Methodology}
\label{sec:methodology}
In this section, we describe the methods used to collect our dataset and the analysis methods.

\begin{figure}[ht!]
    \centering
    \includegraphics[width=\linewidth]{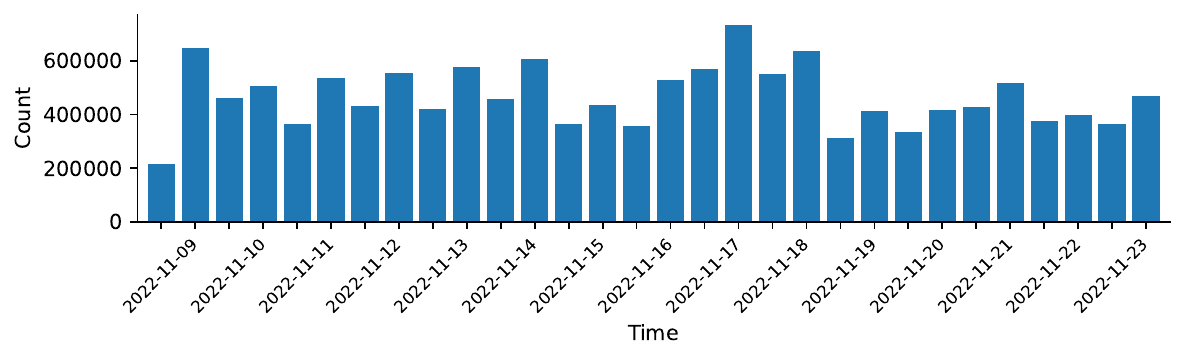}
    \caption{Volume of collected tweets from November 09, 2022, to November 23, 2022, binned in 12-hour intervals.}
    \label{fig:tweet_volume}
\end{figure}

\begin{table}[ht!]
\centering
{\small
\begin{tabular}{|l|l|}
\hline
\multicolumn{1}{|c|}{\textbf{Field}} & \multicolumn{1}{c|}{\textbf{Description}} \\ \hline
id                                   & Unique ID of the tweet                        \\ \hline
timestamp\_ms                        & Timestamp of the tweet                    \\ \hline
user.id                              & User ID                                   \\ \hline
full\_text                           & Main text of the tweet                        \\ \hline
quote\_count                         & Number of quotes on the tweet             \\ \hline
reply\_count                         & Number of replies on the tweet            \\ \hline
retweet\_count                       & Number of retweets on the tweet           \\ \hline
favorite\_count                      & Number of ``favorites'' on the tweet      \\ \hline
quoted\_status.id                    & ID of tweet being quoted                  \\ \hline
quoted\_status.user.id               & User ID of the tweet being quoted            \\ \hline
in\_reply\_to\_status\_id            & ID of the tweet being replied to              \\ \hline
in\_reply\_to\_user\_id            & User ID of the tweet being replied to              \\ \hline
\end{tabular}
}

\caption{Features Used in Analysis}
\label{tab:features_exmple}
\end{table}

\subsection{Data Collection}
We manually curated a list of keywords related to cryptocurrency and security, ranging as broadly as ``crypto'' or ``blockchain'' to specific technologies such as ``zk-rollup'' or ``layer-2.''
We also included security-related terms such as ``breach'' or ``hack'' to get tweets related to cryptocurrency hacks.
In total, we collected our dataset using a list of 199 keywords.
The full list of keywords is available in our anonymized repository.
The data collection ran over 14 days, from November 09, 2022, to November 23, 2022.
The final dataset consists of 39,828,572 unique tweets.

We used Twitter API v1~\cite{twitterapi} to scan and scrape tweets that contain the selected keywords in real-time, maintaining our databases using Elasticsearch~\cite{elasticsearch2018elasticsearch}.
If a new tweet contains one or more keywords, it is added to the Elasticsearch index.
The volume of tweets during our collection period can be seen in \Cref{fig:tweet_volume}.
The complete index contains 130 features about each tweet.
We considered a subset of these features, such as the text or reply/quote structure, in our analysis.
Some of the features used in our work are shown in \Cref{tab:features_exmple}.

\subsection{Data Sanitization}
We found that some of the tweets in the dataset do not discuss any cryptocurrency-related topics, such as discussing just about ``security'' unrelated to the domain we are exploring.
However, in the later stages of our analysis, the cluster-based and graph-based approaches can identify such tweets and group them accordingly.

In the context of Crypto Twitter, a \textit{spam} tweet refers to an unsolicited and often irrelevant message, typically intended to promote certain cryptocurrencies, ICOs, or other crypto-related schemes.
These tweets are usually characterized by their repetitive nature, excessive hashtags, and links to external sites.
They might also include misleading information, false promises of high returns, or attempts to impersonate well-known figures in the crypto community.
The primary goal of spam tweets in the crypto context is often to manipulate market sentiment, spread misinformation, or engage in fraudulent activities such as phishing or scams.
Unsurprisingly, we found many such \textit{spam} tweets present in our initial dataset of 40  million tweets.

However, we also note that some of these tweets could have been made by genuine users who share a lot of enthusiasm about the topic or are after an ``Airdrop" of a certain crypto token~\cite{allen2023airdrop}.
This introduces some difficulties regarding our approach to definitively identifying spam tweets.
With this in mind, we only removed tweets verified to be linked with a scam attempt.
For example, we found that millions of tweets in our dataset contain the same set of sentences ``Uniswap is being exploited by this dude. Why is nobody talking about this?...''.
These tweets appear to have been made by a coordinated bot attempt to push a phishing campaign~\cite{sinclair2022uniswap}, and were thus removed from our dataset.

After manually filtering the tweets by matching commonly found spam patterns that we were able to verify, we ended with 20,911,310 tweets in the final dataset for the analysis.

In addition to filtering for spam tweets, we replace user mentions with \textit{user} and any external links to \textit{http}, which allows us to better capture the style of tweets found in the dataset instead of specific mentions of individual users or links.

\subsection{Sentence Embedding}
We used a publicly available pre-trained Sentence BERT (SBERT) \cite{reimers2019sentence} model from HuggingFace~\cite{huggingface-sbert} to encode the collected tweets into a latent space.
SBERT is a BERT-based model that has been shown to outperform BERT in both \textit{SentEval} and \textit{Semantic Text Similarity (STS)} tasks \cite{reimers2019sentence}.
External URLs and user mentions were replaced with the keywords ``http,'' and ``user,'' respectively, to avoid overfitting on terms that appear in usernames and URLs.

\subsection{Sentiment Analysis}
We used a publicly available RoBERTa~\cite{liu2019roberta} model fine-tuned on the TweetEval sentiment analysis task benchmark~\cite{barbieri2020tweeteval} to generate the sentiment scores used in our analysis.
The publicly available weights from HuggingFace~\cite{twitter-roberta} were used without any additional fine-tuning.
Unlike Sentence-BERT which is fine-tuned using a contrastive objective to predict the semantic similarity of sentences, this RoBERTa model was fine-tuned with a classification layer for predicting sentiment labels corresponding to the input text.
The sentences are fed to the model to yield three types of sentiments [\textit{positive}, \textit{neutral}, and \textit{negative}] and the confidence score associated with each label.

\subsection{Tweet Clustering}
\begin{figure}[ht!]
  \centering
  \includegraphics[width=0.8\linewidth]{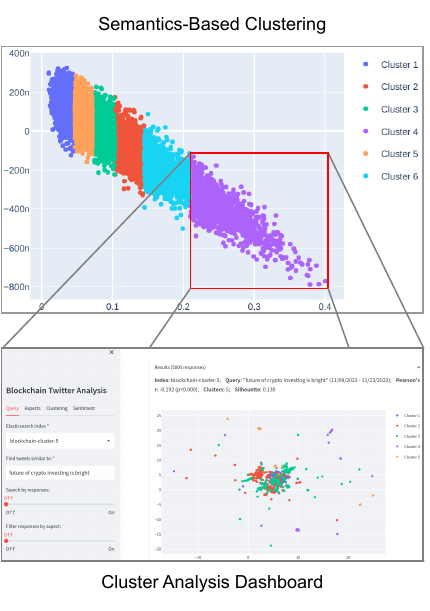}
  \caption{
    Pipeline of the two-stage topic clustering process.
    1) The BERT embeddings are used to form clusters of the entire dataset using K-means.
    2) Each cluster is then examined manually via a web dashboard for further analysis.
  }
  \label{fig:clustering_pipeline}
\end{figure}

In order to cluster the tweets by their semantic similarity, we used the SBERT embeddings~\cite{reimers2019sentence}.
We reduced the dimensionality of the embeddings by training and applying a linear autoencoder, then performed K-means clustering to group the tweets by their similarity in the latent space.
We then used a term-frequency inverse-document-frequency (TF-IDF) matrix to extract the topics for each group.
Even after dividing the tweets into different clusters, we faced a very large number of tweets to examine in each cluster.
Therefore, to aid in this process, we implemented a dashboard (see \Cref{fig:clustering_pipeline})
to search each cluster of tweets effectively.
Given a certain query string, this dashboard embeds the query into a SBERT embedding and searches for tweets with similar embeddings (nearest-neighbors by cosine similarity).
Using the keywords observed in our TF-IDF analysis, we were able to take a deeper dive into the collected tweets by manually building queries using keywords contained in our clusters.

In order to get a cluster-specific set of topics, we excluded the top 10 terms found in the overall dataset from the top terms of each cluster.

We extracted the sentiment score of each tweet's raw text using the RoBERTa model fine-tuned on the TweetEval sentiment benchmark.
To understand the sentiment change over time, we analyzed the keywords associated with each type of sentiment.
We then grouped the tweets in each cluster by their tagged sentiment type and conducted TF-IDF analysis on each sentiment group separately.

\subsection{Network-Based Analysis}

In addition to clustering tweets using a semantics lens, we examined the dataset at the thread and user levels by constructing graphs of tweets.
For this purpose, the \texttt{\small reply\_to} and \texttt{\small quoted\_status} fields of the tweets were utilized that contain the user ID and the tweet ID if it exists.
We chose these fields because we wanted to model the user-to-user and tweet-to-tweet interactions to understand the conversation dynamics better. 
To achieve this, we constructed different graphical views of user interaction and tweets using these fields and analyzed their reply and quote interactions.
We thus constructed and analyzed four types of graphs: tweet-quote, tweet-reply, user-quote, and user-reply.

We considered two types of components to analyze, namely \textit{weakly connected components} (WCC) and \textit{strongly connected components} (SCC).
A weakly connected component is a subgraph where each node is reachable to every other node \textit{disregarding} the direction of the edges.
A strongly connected component is a subgraph where each node is reachable to every other node \textit{while considering} the direction of the edges.
Our analysis involving graphs was implemented using Python networkx library~\cite{networkx} and Gephi visualization~\cite{gephi}.

\section{Analysis Results}

The motivation behind our analysis is twofold. 
We wish to understand the types of topics discussed in Crypto Twitter, as well as the conversation dynamic present in it.
Thus, we performed two different types of analysis on our dataset.
In the first analysis, we grouped the tweets and analyzed the topics and the sentiments associated with each group.
This method focuses on understanding the types of tweets that exist in Crypto Twitter, such as the topics and the associated sentiment values.
The second analysis is graph-based, in which we constructed a tweet-level graph based on the \textit{reply} and \textit{quote} relationships of the tweets as well as a user-level graph with the same \textit{reply} and \textit{quote} relationships. 
This method aims to better understand the dynamics of conversation within Crypto Twitter.

\subsection{Topic Clustering}

\begin{table}[ht!]
{\small
\begin{tabular}{|c|l|p{5cm}|}
\hline
\textbf{ID} & \textbf{Characterization} & \textbf{Top 10 Keywords}                                                      \\ \hline
            & \emph{Overall}            & user, http, pump, just, signal, happen, crypto, wallstreetbets, event, kucoin \\ \hline
1           & Crypto Conspirators       & user, pump, http, signal, just, event, happen, wallstreetbets, kucoin, big    \\ \hline
2           & Meta Crypto Twitter       & user, crypto, http, promote, roll, token, price, 000, binance, security       \\ \hline
3           & Crypto Observers          & user, http, crypto, v2, rollup, address, tokens, claiming, compatible, evm    \\ \hline
4           & Crypto Commenters         & roll, crypto, sushi, user, project, security, good, try, http, bridge         \\ \hline
5           & Crypto Doubters           & user, http, pump, crypto, just, signal, kucoin, happen, event, wallstreetbets \\ \hline
6           & Interested Investors      & promote, user, crypto, price, roll, http, btc, binance, eth, bitcoin          \\ \hline
\end{tabular}
}
\caption{Top 10 Words per Cluster}
\label{tab:top_topic_count}
\end{table}

The keywords associated with positive and negative sentiment in each cluster can be seen in \Cref{tab:top_10_positive} and \Cref{tab:top_10_negative}, respectively.
Similar to how the clusters' topics were found, we excluded the top 10 terms of each sentiment at the dataset level to reduce the number of overlapping terms across clusters.

In order to capture the keywords of each cluster, we fitted a TF-IDF matrix over the entire document (every tweet) and per cluster, then selected the top 10 keywords for the overall document and each cluster.
 This process removes the most frequent keywords at the document level from each cluster, allowing for more granular insight into each cluster.
The document-level keywords are marked as \textit{overall} at the top row of each table.
However, some overlap between the keywords of different clusters can still be observed. These can be interpreted as the overall trend captured in our dataset that our keyword filtering process could not remove.

\begin{figure}
    \centering
    \includegraphics[width=.9\linewidth]{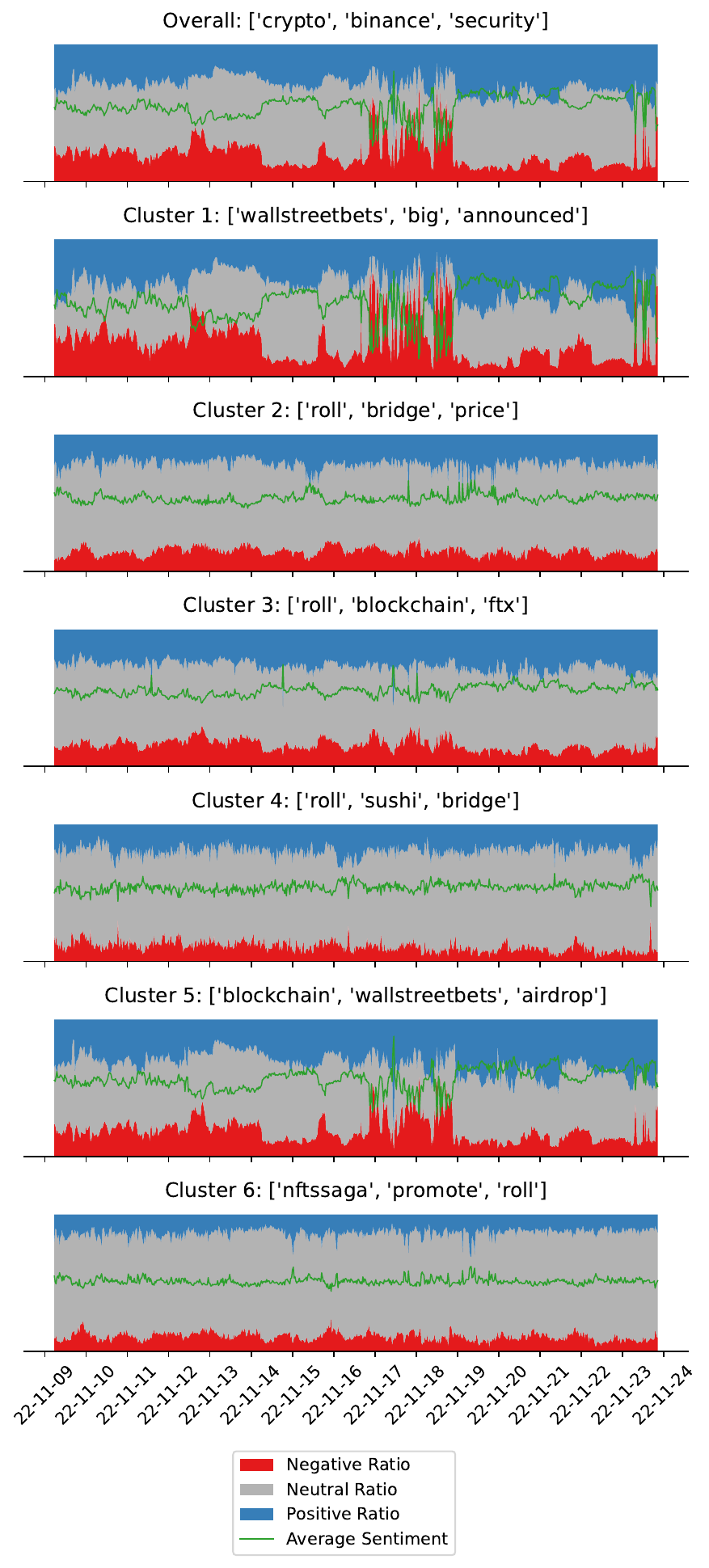}
    \caption{Change in sentiment score over time for each cluster with top-3 cluster-specific terms.\\
    \footnotesize{
    blue (\textcolor{blue}{\textbf{---}}): ratio of positive sentiment,
    red (\textcolor{red}{\textbf{---}}): ratio of negative sentiment,
    gray (\textcolor{gray}{\textbf{---}}): ratio of neutral sentiment,
    green (\textcolor{green}{\textbf{---}}): average sentiment
    }
    }
    \label{fig:sentiment_by_count}
\end{figure}

The overall sentiment change over time in each cluster can be seen in \Cref{fig:sentiment_by_count}.

\bigskip
\noindent\textbf{Cluster 1}: 
Primarily comprises topics related to \emph{group activity} or \emph{price manipulation}.
It contains tweets advertising private price manipulation groups, similar to the phenomenon noted by \cite{nizzoli2020charting}.
The two main targets of these price manipulation tactics are Binance and Kucoin, two popular cryptocurrency exchanges.
An in-depth analysis of characteristic tweets reveals multiple ``pump signals'' announcements on Binance and Kucoin by groups attempting to perpetrate pump-and-dump schemes.
The amount of identical or near-identical tweets makes this a noteworthy example of spam within the dataset relevant to mass crypto-exploitation and potentially harmful trading activity.
This observation is supported by several of the top 10 words per cluster seen in \Cref{tab:top_topic_count}.
%
Top terms associated with positive sentiment are related to announcements of events, such as ``massive,'' ``just,'' or ``announced''.
We also found that ``binance'' appears as a positive term, while ``ftx'' appears as a negative term.
Most of the negative sentiment associated with FTX can be linked to the event where the head of Binance decided to stop their support for FTX~\cite{Sigalos2022Nov}.
We found tweets expressing distrust towards FTX even before the public announcement of the event,
and threads discussing the news of Binance walking away from a deal with FTX, expressing their distrust towards FTX and its founder.
We also found negative sentiment towards not just FTX but also people who supported FTX, with many tweets blaming the victims for trusting FTX.

\begin{table}[h]
\centering
{\small
\begin{tabular}{|c|p{7cm}|}
\hline
\textbf{ID}    & \textbf{Top 10 Keywords}                                                                \\ \hline
\emph{Overall} & crypto, project, binance, token, signal, blockchain, pump, happen, good, big            \\ \hline
1              & wallstreetbets, massive, announced, event, really, airdrop, kucoin, join, best, awesome \\ \hline
2              & roll, love, security, best, great, kinderinu, world, bridge, join, 10                   \\ \hline
3              & roll, best, security, love, amazing, join, great, nft, like, team                       \\ \hline
4              & roll, sushi, nice, want, love, great, dodo, let, bridge, security                       \\ \hline
5              & best, join, wallstreetbets, love, massive, nft, airdrop, awesome, announced, security   \\ \hline
6              & roll, sushi, security, kinderinu, love, new, bridge, nfts, best, things                 \\ \hline
\end{tabular}
}
\caption{Top 10 Terms Associated With Positive Sentiment}
\label{tab:top_10_positive}
\end{table}

\begin{table}[h]
\centering
{\small
\begin{tabular}{|c|p{7cm}|}
\hline
\textbf{ID}    & \textbf{Top 10 Keywords}                                                  \\ \hline
\emph{Overall} & crypto, security, people, roll, like, hack, ftx, binance, bridge, money   \\ \hline
1              & know, uniswap, want, time, blockchain, social, think, going, need, market \\ \hline
2              & know, going, shit, token, threat, want, bad, time, scam, think            \\ \hline
3              & know, want, going, time, think, threat, token, need, bad, national        \\ \hline
4              & sushi, bad, shit, malicious, scam, fuck, losers, dumps, hour, trade       \\ \hline
5              & know, want, going, time, blockchain, think, national, need, social, token \\ \hline
6              & sushi, going, bad, shit, 2022, scam, malicious, know, vulnerability, fuck \\ \hline
\end{tabular}
}
\caption{Top 10 Terms Associated With Negative Sentiment}
\label{tab:top_10_negative}
\end{table}

\bigskip
\noindent\textbf{Cluster 2}: 
Unlike in cluster 1, cluster 2's top 10 keywords were less useful in determining the overall topic of the cluster, because several words, like `security,' `roll,' and `bridge,' are associated with many tweets unrelated to cryptocurrency and DeFi.
For example, using the keyword ``crypto'' did not initially reveal a unifying topic of discussion, as tweets ranged from short statements like \textit{``crypto is the future''} to \textit{``Why should someone use crypto''}.
However, using our two-part clustering method, we found that one sub-cluster of cluster 2 contained many tweets directed toward Elon Musk, who has often weighed in on crypto-related events and promoted various coins.
Since Elon Musk is an influential user within the DeFi space, there were many tweets asking for his opinion on specific projects or requesting promotion for tokens. 
 
We found keywords such as ``project,'' ``blockchain,'' and ``token'' associated with positive sentiment, while keywords such as ``hack,'' ``roll,'' ``bridge'' are associated with negative sentiment.
Many of the positive sentiments were originated from tweets expressing their gratitude towards developers.
We found many tweets similar to ``Great project'' or ``Thank you for this project'' in positive sentiments.
Looking into the negative sentiments, we found tweets expressing their distrust in ``bridge hacks'' and the security features of many bridge protocols.
We also found many specific mentions to the ``Ronin Bridge'', a project related to Axie Infinity, which was hacked in July~\cite{Kaaru2022Jul} with a loss of over 600 million dollars.

\bigskip
\noindent\textbf{Cluster 3}:
Here, we found chatter regarding cryptocurrency exchanges, with users expressing distrust in these institutions and, less commonly, optimism regarding the trajectory of DeFi.
Searches using the keywords ``binance'' and ``FTX'' revealed a general mistrust in exchanges.
Interestingly, mistrust of centralized exchanges (CEXes), such as FTX, does not always translate to mistrust in cryptocurrency as a whole, with many users advocating for a shift away from CEXs toward more decentralized methods of digital currency storage, such as offline wallets.
These opinions are unsurprising in light of the FTX incident~\cite{Sigalos2022Nov}, which occurred during data collection and whose effects are still felt within the market months later.

We found many overlapping terms for positive and negative sentiment.
Terms such as ``security'' and ``roll '' appear in both sentiments, suggesting that multiple sub-clusters in this cluster express their trust and distrust of centralized exchanges regarding these topics.
We also find that ``binance'' appears as one of the top positive terms, while ``ftx'' appears among the negative terms.
Looking deeper, we find that many of the tweets with a positive sentiment that mentions Binance are thanking the CEO of Binance.
In contrast, FTX is mostly mentioned in a negative context, especially after the official bankruptcy announcement.

\bigskip
\noindent\textbf{Cluster 4}:
This cluster contained the most syntactically distinct tweets, with most only two or three words long.
Common tweets within this cluster were \textit{``crypto fan''} and \textit{``to the moon''}, a phrase typically used to signify one's hope that a project will take off and rapidly accrue investors.
These quick comments are in response to other tweets intended to generate engagement, and thus represent a subset of ordinary users within Crypto Twitter space rather than influencers or official project social media accounts.

As pointed out in our earlier analysis, cluster 4 comprises mostly short-form tweets.
Many positive and negative keywords are either adjectives or verbs with strong sentiment attached, such as ``nice,'' ``love,'' ``want'' for positive sentiment, ``malicious,'' ``hack,'' and ``bad'' for negative sentiment.
We also note that the term ``security'' appears in both positive and negative sentiment, suggesting that tweets in this cluster also discuss the security aspect of various projects in both positive and negative contexts.

\bigskip
\noindent\textbf{Cluster 5}: 
This cluster contained tweets related to the security of various blockchain projects, with users tweeting directly at the official accounts of various projects to ask questions regarding security and trust.
This characterization is upheld when querying for ``binance.''
Many tweets discuss transparency and the duties of exchanges to their customers.
Tweets supporting increased transparency and regulations on blockchain trading activities show the community's growing interest in ensuring safety and reliability in projects that handle large amounts of money.
 
In our earlier analysis of the topics, we found that this cluster is the least coherent in terms of the topics and contains many advertisements for different blockchain assets/protocols.
As a result of this, we did not find any specific mentions of projects or exchanges except for Binance, which was a very engaging topic at the time of our data collection.
Our sentiment analysis supports the finding that most positive sentiments are associated with advertisements, with keywords such as ``just,'' ``join,'' and ``best'' appearing in the top positive terms. 

\bigskip
\noindent\textbf{Cluster 6}:
In this cluster, we found a multitude of tweets discussing various tokens.
Notably, this differs from the promotional activity seen in cluster 1.
Unlike the spam advertisements that dominate the advertisement space of that cluster, these tweets are unique affirmations or criticisms by users often tweeted directly at official project accounts.
Querying this cluster using the keyword `token' yielded tweets such as \textit{``I trust in WorldCupinu token"} and \textit{``\$CREAM token still flying.''}
This shows that despite the various scam activities in the crypto space, some users have genuine faith in ongoing projects.

We found that cluster 6 has the highest number of specific names mentioned in the blockchain space associated with positive sentiments, such as ``binance,'' ``kinderinu,'' and ``sushi''.
Sushi is likely in reference to the Sushi Swap decentralized exchange, while Kinderinu (\emph{\url{kinderinu.io}}) is a type of ``meme'' token similar to Dogecoin (\emph{\url{dogecoin.com}}), which gained popularity among some investors.
Looking deeper, we found that many tweets that mention the ``kinderinu'' project have the same content, similar to the Uniswap liquidity spam tweets mentioned in the Methodology section.
Some of the tweets in this cluster were made under a coordinated attempt, although we could not find a source to verify.


\begin{table*}[!ht]
{\small
\begin{tabular}{|l|c|c|ccc|ccc|ccc|}
\hline
\multirow{2}{*}{\textbf{Graph Type}} & \multirow{2}{*}{\textbf{\# Nodes}} & \multirow{2}{*}{\textbf{\# Edges}} & \multicolumn{3}{c|}{\textbf{Degree}}                                           & \multicolumn{3}{c|}{\textbf{In-Degree}}                                       & \multicolumn{3}{c|}{\textbf{Out-Degree}}                                       \\ \cline{4-12} 
                                     &                             &                             & \multicolumn{1}{c|}{Mean\footnotesize{/std}}        & \multicolumn{1}{c|}{Median} & Max  & \multicolumn{1}{c|}{Mean\footnotesize{/std}}        & \multicolumn{1}{c|}{Median} & Max & \multicolumn{1}{c|}{Mean\footnotesize{/std}}        & \multicolumn{1}{c|}{Median} & Max  \\ \hline
Tweet-Quote                          & 44,300                       & 26,255                       & \multicolumn{1}{c|}{1.185\footnotesize{/1.893}} & \multicolumn{1}{c|}{1}      & 184  & \multicolumn{1}{c|}{0.592\footnotesize{/0.491}} & \multicolumn{1}{c|}{1}      & 1   & \multicolumn{1}{c|}{0.592\footnotesize{/1.998}} & \multicolumn{1}{c|}{0}      & 184  \\ \hline
Tweet-Reply                          & 1,591,247                     & 1,034,017                     & \multicolumn{1}{c|}{1.299\footnotesize{/5.952}} & \multicolumn{1}{c|}{1}      & 6,391 & \multicolumn{1}{c|}{0.650\footnotesize{/0.477}} & \multicolumn{1}{c|}{1}      & 1   & \multicolumn{1}{c|}{0.650\footnotesize{/5.978}} & \multicolumn{1}{c|}{0}      & 6,391 \\ \hline
User-Quote                           & 1,640,251                            & 2,520,608                             & \multicolumn{1}{c|}{3.073\footnotesize{/144.905}}            & \multicolumn{1}{c|}{1}       &  59,825    & \multicolumn{1}{c|}{1.537\footnotesize{/6.068}}            & \multicolumn{1}{c|}{1}       & 3,736    & \multicolumn{1}{c|}{1.537\footnotesize{/144.788}}            & \multicolumn{1}{c|}{0}       & 59,825      \\ \hline
User-Reply                           &         2,340,960                    &          7,298,105                   & \multicolumn{1}{c|}{6.235\footnotesize{/75.741}}            & \multicolumn{1}{c|}{1}       &   49,486   & \multicolumn{1}{c|}{3.118\footnotesize{/52.859}}            & \multicolumn{1}{c|}{1}       & 13,051     & \multicolumn{1}{c|}{3.118\footnotesize{/52.453}}            & \multicolumn{1}{c|}{1}       & 49,481      \\ \hline
\end{tabular}
}
\caption{Statistics of the Tweet and User Reply/Quote Graphs}
\label{tab:graph-stats}
\end{table*}


\subsection{Tweet Graph}

The tweet graph is a directed graph in which each node is a tweet.
A directed edge from tweet A to tweet B indicates that tweet A is a reply or quote to tweet B.
We filtered for tweets that contained valid text to gather a subset of the dataset,  which excluded retweets and tweets that refer to another tweet with a quote/reply. 
Consequently, each node in our graph is associated with its respective text content.
Additionally, we computed RoBERTa sentiment scores for these texts. Given that each tweet (node) can quote or reply to at most one other tweet, resulting in at most one outgoing edge per node, our tweet graph tends to be sparse. 
Therefore, our analysis primarily concentrated on WCCs, as these represent more significant clusters or groups within the reply and quote dynamics of the graph.


\begin{figure}[!ht]
    \centering
    \begin{tikzpicture}
        \node[anchor=north] (graph) at (0,0) {\includegraphics[width=.5\linewidth]{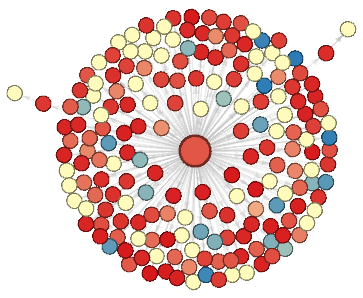}};
        \node[inner sep=0pt] (colorbar) at ($(graph.south)+(0,-5pt)$) {\includegraphics[width=.4\linewidth]{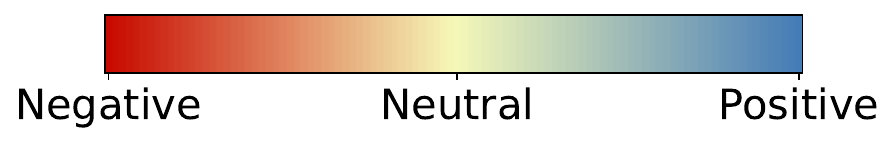}};
    \end{tikzpicture}
    \caption{
    Largest WCC of the Tweet-Quote graph.\\
    $|V| = 188, |E| = 187$.
    }
    \label{fig:tweet-quote-wcc-0}
\end{figure}

\begin{figure}[!ht]
    \centering
    \begin{tikzpicture}
        \node[anchor=north] (graph) at (0,0) {\includegraphics[width=.7\linewidth]{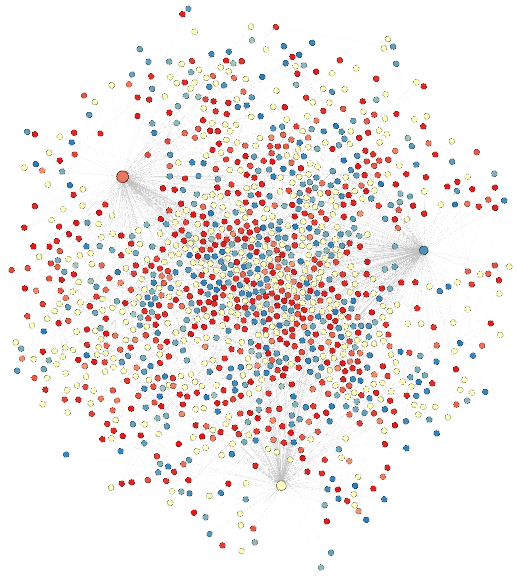}};
        \node[inner sep=0pt] (colorbar) at ($(graph.south)+(0,-5pt)$) {\includegraphics[width=.4\linewidth]{figures/Results/graph/graph_colormap.pdf}};
    \end{tikzpicture}
    \caption{
    Seventh largest WCC of the Tweet-Reply graph.\\
    $|V| = 1282, |E| = 1281$.
    }
    \label{fig:tweet-reply-wcc-6}
\end{figure}

The statistics of the generated graph can be seen in~\Cref{tab:graph-stats}.
As seen in the degree distribution, the quoted graph is much sparser than the reply graph.
This indicates that users in our dataset communicate more frequently using replies instead of quotes.
We identified the ten largest WCCs for each version of the node relationship considered and analyzed the types of tweets present in each of the largest WCCs.
Examples of the tweet-reply and tweet-quote graphs can be seen in~\Cref{fig:tweet-quote-wcc-0} and~\Cref{fig:tweet-reply-wcc-6}.
The nodes are colored by their sentiment and sized by their interaction count.
\begin{itemize}
    \item Nodes with shades of red (\textcolor{red}{$\bullet$}) have negative RoBERTa sentiment, darker color being stronger.
    \item Nodes with yellow (\textcolor{betteryellow}{$\bullet$}) have neutral RoBERTa sentiment.
    \item Nodes with shades of blue (\textcolor{blue}{$\bullet$}) have positive RoBERTa sentiment, darker color being stronger.
    \item Nodes are sized by the \textit{interaction count}, which is a sum of \texttt{\small retweet\_count}, \texttt{\small favorite\_count}, \texttt{\small reply\_count} and \texttt{\small quote\_count}.
\end{itemize}

\noindent\textbf{Tweet-Quote Graph}:
The tweets in the largest WCC (WCC-1) discuss an incident from Crypto.com, a centralized cryptocurrency exchange, in which the exchange had accidentally misplaced its funds~\cite{sarkarCryptoComAccidentally2022}.
The interactions appear from genuine users, many expressing their confusion and concerns about the incident.
A snapshot of this interaction can be seen in~\Cref{fig:tweet-quote-wcc-0}.
The central tweet in the sixth largest WCC (WCC-6) discusses that some cryptocurrency exchanges appear to be swapping funds to build a fake \textit{snapshot} of their reserves. 
Further inspection reveals that this tweet was made by a Crypto Twitter influencer, referencing a scandal where several centralized exchanges were suspected to lack the liquidity they claim~\cite{sarkarHuobiGateIo2022}.
All the other WCCs (WCC-2, WCC-3, WCC-4, WCC-5, WCC-7, WCC-8, WCC-9, and WCC-10) contain mostly bot attempts, and a few notable bot activities can be characterized as follows:
\begin{itemize}
    \item Tweets attempting to shine a positive light on the Terra-Luna project after its collapse in May~\cite{liu2023anatomy}.
    \item Fake airdrop scams that advertise a token from a project that had not been released or advertised, leading to a third-party website for phishing~\cite{agboAirdropScamsCrypto}. For instance, our dataset had many mentions of zksync (\emph{\url{zksync.io}}), a platform known for its airdrops, and the users linking a URL to a different (phishing) website.
\end{itemize}

\noindent\textbf{Tweet-Reply Graph}:
The largest WCC (WCC-1) in this graph mostly comprises tweets with the same two hashtags about a token called ``Leonicorn.''
At first glance, this appears to be a group of spam tweets. 
However, many of the user accounts behind these tweets appear legitimate, with up to hundreds of followers and previous activity on Twitter.
We conclude that this may be an example of a coordinated attempt by users to bring certain hashtags to the \textit{trending} list of hashtags on Twitter.
While the accounts appear to be regular Twitter users, this activity reveals another example of manipulation on Crypto Twitter.

The third, fourth, and tenth largest WCCs (WCC-3, WCC-4, WCC-10) reveal another interesting insight into spam activity on Crypto Twitter.
There is a reply chain composed almost entirely of spam bots with regular users scattered throughout.
The bot accounts appear to reply to seemingly randomly chosen users who discuss cryptocurrency-related topics.
We also found instances where bots advertising different scams reply to each other in a chain that starts with a tweet from a genuine user.
Many of these bots advertise an ``exploit'' the user can benefit from, such as Maximal Extractable Value (MEV) scripts, which exploit the ability to influence the order of transactions in a block to gain advantages, or an upcoming ``strategy'' such as a rug pull on an asset, where the developers of a cryptocurrency project suddenly withdraw all their funds from the liquidity pool or project wallet, abandoning the project and leaving investors with worthless tokens or assets.
We found several instances where regular users attempt to respond to the bot tweets, only to be swarmed by further tweets advertising a different ``exploit.''

The sixth largest WCC (WCC-6) is centered around a tweet announcing a giveaway event from Binance's official account. 
While this tweet is legitimate, most replies are tweets from low-follower users that link to an external website in a likely scam attempt, which reveals another pattern of spam bots on Crypto Twitter.
Unlike other spam tweets observed in the dataset, these tweets only mention the official Binance account.
Instead of mentioning multiple users with a description of the ``exploit,'' these tweets use a few words such as \textit{``Interesting''} or \textit{``Important.''}
We conclude that these tweets attempt to lead possible victims into visiting the link out of curiosity and then fall for the scam.
Furthermore, we found that many of these links lead to deleted videos on YouTube
, which was a popular format for sharing the ``tutorials'' on these ``exploit'' scams.

The second and fifth largest WCCs (WCC-2, WCC-5) appear to stem from Binance's founder and former CEO Changpeng Zhao. 
WCC-2 is centered around a tweet from Zhao, which says \textit{``I think I am a poor speaker. I stutter all the time.''}
This tweet is met by a majority of positive support from the community, with the users cheering him on and expressing their faith in him.
WCC-5 is centered around Zhao's tweet \textit{``Rebuilding starts, to the moon.''} announcing a rebuilding of the crypto community after the FTX incident.
This tweet is met with mixed sentiment, with many users expressing distrust in the centralized exchange model. 

Interestingly, when Binance's official account tweets, those are met with much more negative responses.
The seventh largest WCC (WCC-7) contains three tweets from Binance's official account, which appear to be from a thread the account posted on its position in light of the FTX instance.
Even though the tweets attempt to explain Binance's decision not to support FTX, many tweets express the users' discontent with Binance's actions and distrust of the platform.
\Cref{fig:tweet-reply-wcc-6} shows a visualization of WCC-7, where the tweets from Binance's official account are the red (\textcolor{red}{$\bullet$}), blue (\textcolor{blue}{$\bullet$}), and yellow (\textcolor{betteryellow}{$\bullet$}) larger nodes in the top left, center right, and center bottom of the diagram, respectively.

A tweet from Etherum's founder, Vitalik Buterin, is found to be the central node of the ninth largest WCC (WCC-9).
Interstingly, while Vitalik's tweet mentions his doubts about the future of ZK-rollup, an overwhelming majority of the replies to this tweet are positive.
We found that many of these replies are not related to the original tweet by Vitalik but are either promoting a likely scam or proposing ideas to him, many of which are classified as having a positive sentiment.

\subsection{User Graph}

We constructed two distinct user graphs: one based on \emph{quotes} and the other on \emph{replies}.
These are also directed graphs, with nodes representing individual users and edges representing interaction between them.
In the user-quote graph, a directed edge from user A to user B signifies that user B has quoted a tweet from user A.
In the user-reply graph, a directed edge from user A to user B indicates that user B replied to a tweet from user A.
Consequently, we filtered and created edges from only those instances that provide both.
Unlike the tweet graph, some nodes in these graphs feature both inbound and outbound edges, along with self-loops.
Thus, we turned our attention to the SCCs to analyze significant groups within both the reply and quote graphs.
This analysis reveals diverse discussions, spam bot networks, and engagement levels in the conversations.

\Cref{tab:graph-stats} presents statistics for the generated graphs. 
Like the tweet graph, the user-reply graph exhibits higher density than the user-quote graph, suggesting a preference among users for replies as their interaction medium over quotes.
We examined the five largest SCCs for both of the graphs.

\begin{table*}[t]
    \centering
    {\small
    \begin{tabular}{|c|c|c|c|c|c|c|c|c|}
        \hline
        \multirow{2}{*}{\textbf{Observation}} & \multicolumn{2}{c|}{\textbf{Mean of Friends}} & \multicolumn{2}{c|}{\textbf{Median of Friends}} & \multicolumn{2}{c|}{\textbf{Mean of Followers}} & \multicolumn{2}{c|}{\textbf{Median of Followers}} \\
        \cline{2-9}
        & \textbf{Within} & \textbf{Reachable} & \textbf{Within} & \textbf{Reachable} & \textbf{Within} & \textbf{Reachable} & \textbf{Within} & \textbf{Reachable} \\
        \hline
        SCC-1 & 1375 & 1202 & 691 & 368 & 90352 & 7429 & 1019 & 198 \\
        \hline
        SCC-2 & 422 & 404 & 207 & 290 & 365 & 271 & 234 & 107 \\
        \hline
        SCC-3 & 3474 & 96 & 420 & 3 & 4080 & 81 & 84 & 27\\
        \hline
        SCC-4 & 2973 & 313 & 2226 & 4 & 17367 & 2558 & 11701 & 8 \\
        \hline
        SCC-5 & 180 & 95 & 158 & 31 & 144 & 71 & 126 & 34 \\
        \hline
    \end{tabular}
    }
    \captionsetup{width=0.8\textwidth}
    \caption{Observations of Mean and Median (rounded to the nearest integer) Friends/Followers of the Nodes Inside SCCs and the Nodes that are Reachable from SCCs (User-Quote Graph)}
    \label{tab:user_quote_fnf_stats}
\end{table*}

\noindent\textbf{User-Quote Graph}:
In the context of SCC-1, the largest component, discussions center around the event regarding the collapse of FTX~\cite{reiff2022the}.
This subnetwork is dominated by legitimate users, as evidenced by the high mean and median number of friends and followers (see~\Cref{tab:user_quote_fnf_stats}).

SCC-2, the second-largest component, is smaller, with 25 nodes and 105 nodes reachable from it.
Intriguingly, nearly all nodes, including reachable ones, are identified as spam bot users responsible for disseminating over 16,000 spam tweets, primarily glorifying AdsCoin (\emph{\url{https://adsexchange.io}}).
Initial manual examination of tweets confirmed their spam bot nature, and subsequent analyses discussed in later sections further support this claim.

The third-largest component (SCC-3), encompassing 21 nodes with 28 reachable nodes, unexpectedly diverges from crypto-related discussions.
The dialogue here revolves around the ``Save Kashmir movement'' in India.
However, since our dataset is filtered with keywords such as attacks, security, etc., relevant to crypto and non-crypto contexts, these data inadvertently entered the dataset, albeit unrelated to the crypto domain.

On a smaller scale, SCC-4, the fourth-largest component, comprising only 11 nodes, has a relatively larger number of reachable nodes (358).
In this subnetwork, individuals associated with Cosmos (\emph{\url{cosmos.network}}), a cryptocurrency powering a network of blockchains that can communicate with each other, engage in diverse topics.
Unlike a singular discussion focus, Cosmos serves as the unifying element bringing nodes together in this subnetwork.

SCC-5, the smallest component with seven and four reachable nodes, lacks a specific central theme.
Instead, it presents a microcosm of a social network where a small group of connected individuals interact on a range of random topics by quoting each other's tweets.


\begin{figure}
    \centering
    \includegraphics[width=\linewidth]{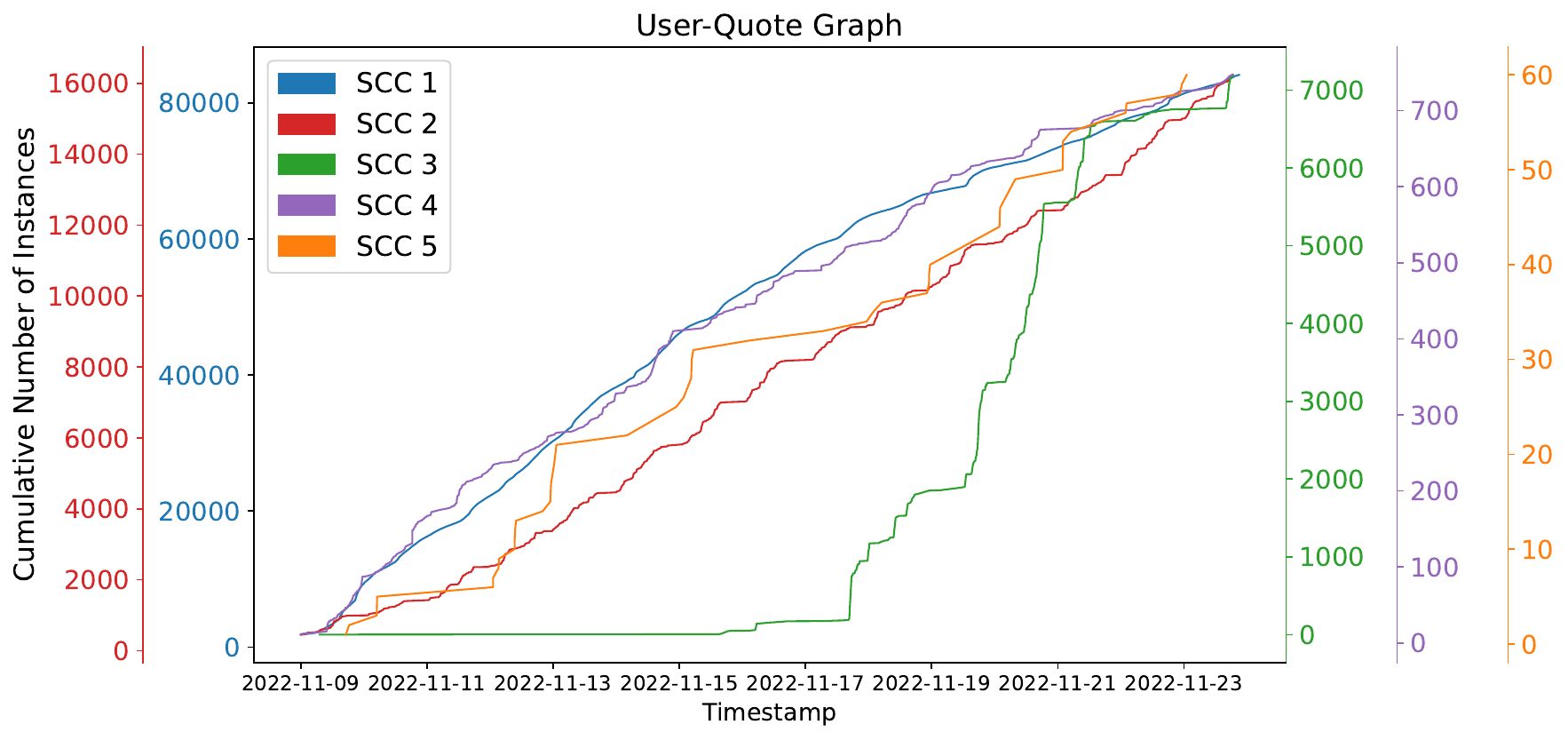}
    \caption{Change in Number of Data Points Involved with the SCC Nodes Over Time (User-Quote Graph)}
    \label{fig:user_quote_scc_evolution}
\end{figure}  

\begin{figure}
    \centering
    \includegraphics[width=\linewidth]{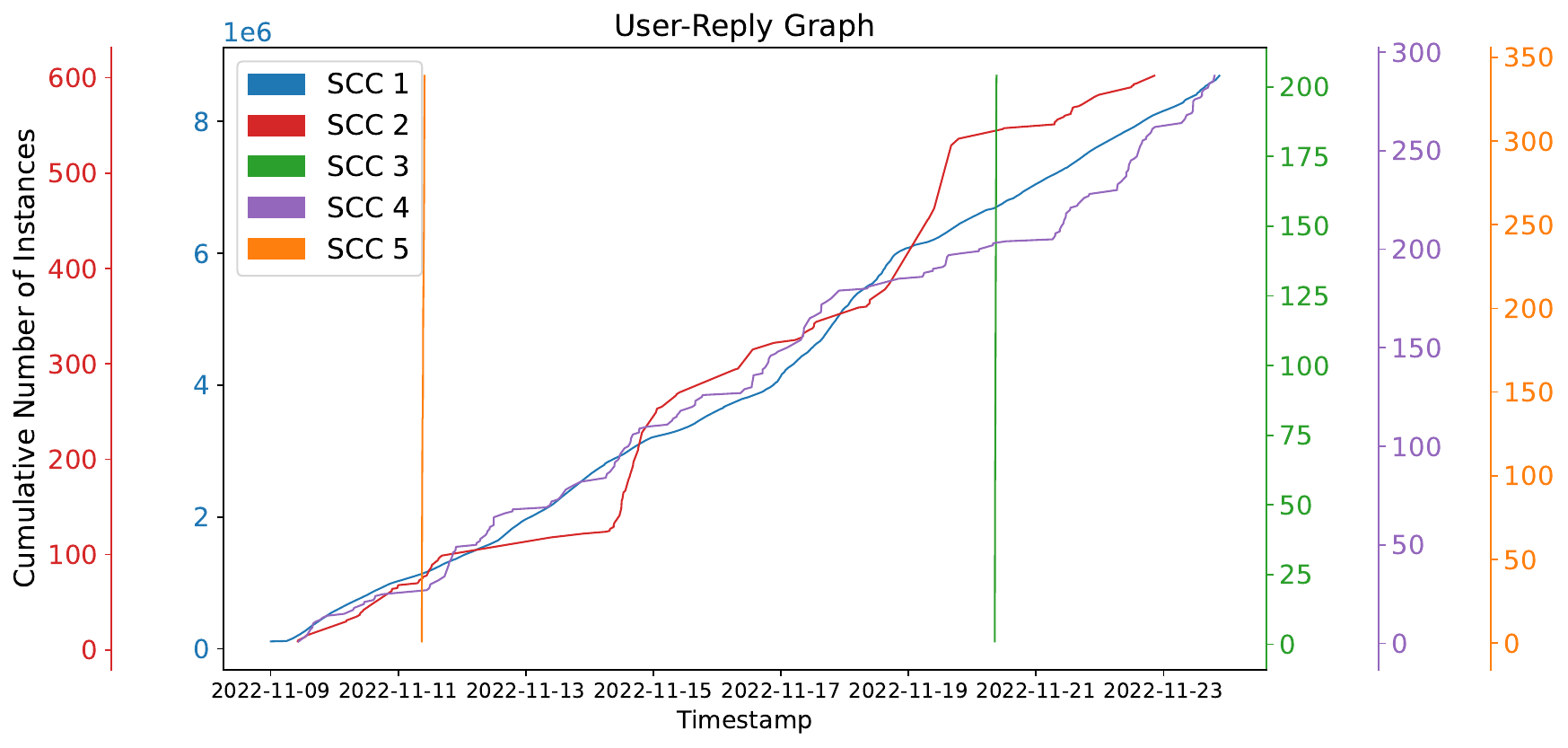}
    \caption{Change in Number of Data Points Involved with the SCC Nodes Over Time (User-Reply Graph)}
    \label{fig:user_reply_scc_evolution}
\end{figure}  

\begin{table*}[t]
    \centering
    {\small
    \begin{tabular}{|c|c|c|c|c|c|c|c|c|}
        \hline
        \multirow{2}{*}{\textbf{Observation}} & \multicolumn{2}{c|}{\textbf{Mean of Friends}} & \multicolumn{2}{c|}{\textbf{Median of Friends}} & \multicolumn{2}{c|}{\textbf{Mean of Followers}} & \multicolumn{2}{c|}{\textbf{Median of Followers}} \\
        \cline{2-9}
        & \textbf{Within} & \textbf{Reachable} & \textbf{Within} & \textbf{Reachable} & \textbf{Within} & \textbf{Reachable} & \textbf{Within} & \textbf{Reachable} \\
        \hline
        SCC-1 & 448 & 296 & 39 & 20 & 26823 & 962 & 4 & 2 \\
        \hline
        SCC-2 & 661 & 436 & 299 & 36 & 5875 & 24535 & 124 & 4 \\
        \hline
        SCC-3 & 0 & 0 & 0 & 0 & 0 & 0 & 0 & 0\\
        \hline
        SCC-4 & 384 & 436 & 63 & 36 & 22487 & 24536 & 33804 & 4 \\
        \hline
        SCC-5 & 180 & 95 & 158 & 31 & 144 & 71 & 126 & 34 \\
        \hline
    \end{tabular}
    }
    \captionsetup{width=0.8\textwidth}
    \caption{Observations of Mean and Median (rounded to the nearest integer) Friends/Followers of the Nodes Inside SCCs and the Nodes that are Reachable from SCCs (User-Reply Graph)}
    \label{tab:user_reply_fnf_stats}
\end{table*}

\noindent\textbf{User-Reply Graph}:
The largest Strongly Connected Component (SCC-1) within the user-reply graph, comprising a substantial number of nodes (231,841), is centered around Binance (\emph{\url{www.binance.com}}), the world's largest cryptocurrency exchange by trading volume.
The majority of accounts within this component are associated with Binance, engaging in discussions on various topics, such as the decision not to pursue the potential acquisition of FTX and attending the Fintech summit in Indonesia.
Although this subgraph is not entirely free from bots, the predominant user base consists of legitimate participants discussing diverse aspects primarily related to Binance.

Interestingly, the second and fourth largest SCCs (SCC-2 and SCC-4) focused on a power outage in Nigeria in November 2022.
Tweets with the highest engagement in SCC-2 originate from the official account of Ikeja Electric (\emph{\url{www.ikejaelectric.com}}), Nigeria's largest Electricity Distribution Company, providing updates and services during the outage.
Similarly, in SCC-4, discussions revolve around another company, Abuja Electricity Distribution Company (\emph{\url{www.abujaelectricity.com}}).
Like the discussions involving SCC-2, these are also focused on updates and services during that power outage.
Although these discussions are not directly related to crypto, Ikeja Electric does mention different tokens, such as energy tokens, which are relevant to their business. 
Notably, the mean followers count for nodes within SCC-2 is significantly lower than that of nodes reachable from SCC-2, deviating from the expected trend (see ~\Cref{tab:user_reply_fnf_stats}).
Further investigation revealed that Elon Musk, who has a very large number of followers, is reachable from SCC-2.
A more in-depth examination reveals the involvement of a spam bot connecting users discussing the power outage and those related to crypto.
Elon Musk replied to a tweet from Chainlink, and a user connected with Chainlink interacted with the spam bot.
This way, Elon Musk entered this network, and his substantial follower base caused the anomaly in the mean followers count.
Other than this deviation, the subnetwork appears normal, although the context in focus is not very related to crypto.
However, SCC-4 doesn't show such deviation and follows the normal trend.

The third largest SCC (SCC-3) is found to be centered around a Binance bot that posted numerous tweets about the CR7 NFT Giveaway in a short timeframe.
This is the same bot network as one of the identified botnets in the tweet-reply graph.

The fifth-largest SCC (SCC-5) also constitutes a spam bot network, where users predominantly post hashtags and random spam replies.
Users here were active for a very short period and posted the tweets and replies in just one day.
Manual verification of users' tweet and reply activities in this subnetwork confirmed their bot nature.
It is worth noting that Twitter has suspended most of these users due to their spamming activities.

The temporal dynamics of user interactions within the SCCs shed light on the evolving nature of discussions over time.
For each of the top 5 SCCs, we observed how the number of data points involved with the SCC nodes changes. To be more specific, for each of the top 5 SCCs, at each timestamp $T$, we counted the number of samples having timestamps less than or equal to $T$ that are connected to/from any of the SCC nodes.

~\Cref{fig:user_quote_scc_evolution} and ~\Cref{fig:user_reply_scc_evolution} illustrate tweet counts over distinct timestamps, thus offering a comprehensive view of the evolution of conversations.
While most SCCs of the user-quote graph exhibit a non-linear pattern that mirrors the dynamic nature of real-life discussions with peaks and lulls, SCC-2 (the red (\textcolor{red}{\textbf{---}}) curve) deviates by showcasing a distinctive linear trend.
This synchronous and unusual linearity aligns with the identified characteristics of SCC-2 as a network primarily composed of spam bot users.
SCC-2 and SCC-5 of the user-reply graph also show distinctive patterns for spam bot networks.
The temporal analysis captures the rhythm of discussions and provides a visual cue to discern patterns that distinguish bot-driven activity from organic, user-driven discourse.

Examining the number of friends and followers within the SCCs and nodes reachable from them accentuates the distinctions (see \Cref{tab:user_quote_fnf_stats,tab:user_reply_fnf_stats}).
The friends and followers counts within most SCCs of the user-quote graph significantly exceed those of reachable nodes.
Notably, SCC-2 follows a distinct trend where the counts within the SCC and the reachable nodes are small and comparable, with the reachable nodes even exhibiting a higher median friends count than the SCCs.
This divergence provides additional evidence supporting the characterization of SCC-2 as a bot network.
These particular statistics in the user-reply graph also follow a similar pattern.
However, SCC-3 with 0 as all the counts indicate an anomaly supporting the previous claim that this subnetwork consists of bots.
The anomalous small counts in SCC-5 and insignificant difference in the counts between the nodes within SCC and the nodes reachable from SCC indicate that this subnetwork is also a spam botnet, reinforcing the previously discussed insight about SCC-5.
These friends and followers count within the SCCs and those reachable from them also reveal a concentration of interaction.
In general, the counts within an SCC surpass those of reachable nodes, indicating that the primary discourse occurs among influential users.
However, the reachable nodes are not negligible users; they serve as observers, sharing tweets through quote tweeting.
More followers than friends signifies the natural trend for influential users.

\section{Conclusion}

We analyzed cryptocurrency-related tweets that we collected during an eventful period in the crypto world (the FTX event) using popular natural language processing techniques such as sentence embedding~\cite{reimers2019sentence}, topic modeling~\cite{grootendorst2022bertopic}, sentiment analysis~\cite{barbieri2020tweeteval}, and by building tweet-level and user-level graphs. 
In our cluster analysis, we successfully identified three types of tweets focused on specific aspects of cryptocurrency, such as investment, security, and price manipulation.
Additionally, our analysis uncovered a significant presence of content creators and developers within these clusters, actively engaging in efforts to generate awareness and excitement for their projects. This engagement pattern offers a valuable resource for understanding user interactions and overlaps across various crypto projects.
%
We also found that a correlation exists between real-life events in the crypto world and the sentiment captured through Twitter data, and that we can estimate the timeline of the FTX incident~\cite{Sigalos2022Nov} using the sentiment spikes on some of our identified clusters.
Interestingly, only some clusters in our dataset displayed a noticeable reaction to the incident, while other topics overshadowed the sentiment in other clusters.
Although our dataset was gathered with a primary focus on cryptocurrency, our clustering results suggest that a more fine-tuned monitoring of social media chatter, such as focusing on one specific project, could reveal more nuances in social media communities' sentiment towards them and offer valuable insights.

Our initial goal was to build a framework for real-time detection of incidents in crypto-twitter.
However, evaluating a real-time detection framework is a difficult task due to the sheer volume of tweets in the space and the unpredictability of such incidents.
Thus, we conducted analyses focusing on fraudulent activity on crypto-twitter, collected around the timeframe of the FTX incident.
Given enough storage and computing resources, our work can be extended as a building block for such a real-time detection system.

While we initially identified that \textit{at least} 8\% of our data are spam tweets, there is a much higher bot activity than we could capture with a simple heuristic spam filter.
Furthermore, our graph-based analysis shed light on the extent of bot activity on Crypto Twitter.
A significant presence of bots was observed at the tweet and user levels, often clustering around popular threads.
This prevalence of bots, coupled with the discovery of legitimate-looking accounts engaged in repetitive tweeting, underscores the challenge of identifying and managing spam content in Crypto Twitter.
However, the clustering algorithm and graph components appear to place these bot tweets in a similar cluster, which allowed us to bypass most of the remaining spam tweets in our dataset.

Our research insights enhance the understanding of the social dynamics of the intricate landscape of Crypto Twitter, revealing the interplay between human and automated interactions. These insights also pave the way for future explorations into the digital ecosystems of cryptocurrency, highlighting the critical need for advanced, nuanced analytical tools to navigate and interpret the ever-evolving narratives within these vibrant online communities.

\bigskip

\noindent \textbf{Resources:} 
All the code used in our analysis is available online in our GitHub repository at~\url{https://github.com/blockchain-interoperability/blockchain-social-media}. 
This includes the Tweet IDs of our dataset, our analysis workflow, and comprehensive documentation.



\section*{Acknowledgements}
The authors acknowledge the support from NSF IUCRC CRAFT Center research grant (CRAFT Grant \#22008) for this research. The opinions expressed in this publication do not necessarily represent the views of NSF IUCRC CRAFT.

\bibliographystyle{ACM-Reference-Format}
\bibliography{references}

\end{document}